\documentclass[aps,prd,10pt,twocolumn,superscriptaddress,showpacs,longbibliography]{revtex4-1}

\pdfoutput=1
\usepackage{hyperref}
\usepackage{graphicx}
\usepackage{amsfonts,amsmath,amssymb,bm,bbm}
\usepackage{color}
\usepackage{enumitem}
\usepackage{url}
\usepackage{multirow}
\usepackage{hhline}
\usepackage[version=4]{mhchem}
\usepackage{slashed}

\hypersetup{
    pdfnewwindow=true,      
    colorlinks=true,       
    linkcolor=blue,          
    citecolor=blue,        
    filecolor=blue,      
    urlcolor=blue        
}

\begin{document}

\title{$\textbf{\textit{W}}$-boson and trident production in TeV--PeV neutrino observatories}

\author{Bei Zhou}
\email{zhou.1877@osu.edu}
\thanks{\scriptsize \!\!  \href{https://orcid.org/0000-0003-1600-8835}{orcid.org/0000-0003-1600-8835}}
\affiliation{Center for Cosmology and AstroParticle Physics (CCAPP), Ohio State University, Columbus, OH 43210}
\affiliation{Department of Physics, Ohio State University, Columbus, OH 43210}

\author{John F. Beacom}
\email{beacom.7@osu.edu}
\thanks{\scriptsize \!\!  \href{http://orcid.org/0000-0002-0005-2631}{orcid.org/0000-0002-0005-2631}}
\affiliation{Center for Cosmology and AstroParticle Physics (CCAPP), Ohio State University, Columbus, OH 43210}
\affiliation{Department of Physics, Ohio State University, Columbus, OH 43210}
\affiliation{Department of Astronomy, Ohio State University, Columbus, OH 43210}

\date{February 18, 2020}

\begin{abstract}
Detecting TeV--PeV cosmic neutrinos provides crucial tests of neutrino physics and astrophysics.  The statistics of IceCube and the larger proposed IceCube-Gen2 demand calculations of neutrino-nucleus interactions subdominant to deep-inelastic scattering, which is mediated by weak-boson couplings to nuclei.  The largest such interactions are $W$-boson and trident production, which are mediated instead through photon couplings to nuclei.  In a companion paper~\cite{Zhou:2019vxt}, we make the most comprehensive and precise calculations of those interactions at high energies.  In this paper, we study their phenomenological consequences.  We find that:
{\bf (1)} These interactions are dominated by the production of on-shell $W$-bosons, which carry most of the neutrino energy,
{\bf (2)} The cross section on water/iron can be as large as 7.5\%/14\% that of charged-current deep-inelastic scattering, much larger than the quoted uncertainty on the latter,
{\bf (3)} Attenuation in Earth is increased by as much as 15\%,
{\bf (4)} $W$-boson production on nuclei exceeds that through the Glashow resonance on electrons by a factor of $\simeq 20$ for the best-fit IceCube spectrum,
{\bf (5)} The primary signals are showers that will significantly affect the detection rate in IceCube-Gen2; a small fraction of events give unique signatures that may be detected sooner.
\end{abstract}

\maketitle


\section{Introduction}
\label{sec_introduction}

The recent detections of TeV--PeV neutrinos by IceCube~\cite{Aartsen:2013jdh, Aartsen:2015knd, Kopper:2017zzm, Stachurska:2019srh, Schneider:2019ayi, Stachurska:2019wfb} are a breakthrough in neutrino astrophysics.  Though the sources of the diffuse flux have not been identified, important constraints on their properties have been determined~\cite{Kistler:2006hp, Beacom:2007yu, Murase:2010cu, Murase:2013ffa, Kistler:2013my, Murase:2013rfa, Ahlers:2014ioa, Tamborra:2014xia, Murase:2014foa, Gauld:2015kvh, Bechtol:2015uqb, Kistler:2016ask, Bartos:2016wud, Bartos:2018jco, Bustamante:2019sdb}.  In addition, there is a candidate source detection in association with a blazar flare~\cite{IceCube:2018cha, IceCube:2018dnn}.  The IceCube detections are also a breakthrough in neutrino physics.  By comparing the observed spectra of events that have traveled through Earth or not, the cross section can be measured at energies far above the reach of laboratory experiments~\cite{Hooper:2002yq, Borriello:2007cs, Connolly:2011vc, Klein:2013xoa, Aartsen:2017kpd, Bustamante:2017xuy}.  And many models of new physics have been powerfully limited by the IceCube data~\cite{Cornet:2001gy, Lipari:2001ds, AlvarezMuniz:2002ga, Beacom:2002vi, Han:2003ru, Han:2004kq, Hooper:2004xr, Hooper:2005jp, GonzalezGarcia:2005xw, Ng:2014pca, Ioka:2014kca, Bustamante:2016ciw, Salvado:2016uqu, Coloma:2017ppo}.

With new detectors --- KM3NeT~\cite{Adrian-Martinez:2016fdl}, Baikal-GVD~\cite{Avrorin:2013uyc}, and especially the proposed IceCube-Gen2 (about 10 times bigger than IceCube)~\cite{Blaufuss:2015muc} --- the discovery prospects will be greatly increased, due to improvements in statistics, energy range, and flavor information.  At high energies, neutrino-nucleus interactions are dominated by deep inelastic scattering (DIS) mediated by weak-boson couplings to nuclei~\cite{Gandhi:1995tf, Gandhi:1998ri}.  For charged-current (CC) interactions, $\nu_e$ leads to a shower, $\nu_\mu$ leads to a shower and a long muon track, and $\nu_\tau$ leads to two showers that begin to separate spatially at $\sim 100$~TeV~\cite{Stachurska:2019wfb}.  For neutral-current (NC) interactions of all flavors, showers are produced.  Cherenkov light is produced by muon tracks and through the production of numerous low-energy electrons and positrons in showers.

With these coming improved detection prospects, new questions can be asked, including the role of subdominant interactions.  We focus on those in which the coupling to the nucleus and its constituents is through a virtual photon, $\gamma^*$, instead of a weak boson~\cite{Lee:1961jj, Czyz:1964zz, Lovseth:1971vv, Fujikawa:1971nx, Brown:1971qr, Brown:1971xk, Brown:1973ih, Belusevic:1987cw, Seckel:1997kk, Alikhanov:2014uja, Alikhanov:2015kla, Altmannshofer:2014pba, Magill:2016hgc, Ge:2017poy, Ballett:2018uuc, Altmannshofer:2019zhy, Gauld:2019pgt}.  The most important processes are on-shell $W$-boson production, in which the underlying interaction is $\nu_\ell + \gamma^\ast \rightarrow \ell + W$, and trident production, in which it is $\nu + \gamma^\ast \rightarrow \nu + \ell_1^- + \ell_2^+$.

In a companion paper~\cite{Zhou:2019vxt}, we make the most comprehensive and precise calculations of these cross sections at high energies.  The cross section of $W$-boson production can be as large as 7.5\% of the DIS cross section for water/ice targets (and as large as 14\% for iron targets, relevant for neutrino propagation through Earth's core)~\cite{Zhou:2019vxt}. For trident production, the most important channels are a subset of $W$-boson production followed by leptonic decays~\cite{Zhou:2019vxt}. To set a scale, IceCube has identified 60 events above 60 TeV in 7.5 years of operation~\cite{Stachurska:2019srh, Schneider:2019ayi}, so taking these subdominant processes into account will be essential for IceCube-Gen2.  Moreover, the $W$-boson and trident events have complex final states, which may allow their detection even sooner, in IceCube.

In this paper, we detail the phenomenological consequences of these processes. In Sec.~\ref{sec_xsec}, we focus on their cross sections.  In Sec.~\ref{sec_detect}, we focus on their detectability. We conclude in Sec.~\ref{sec_concl}.

\begin{figure}[t!]
\includegraphics[width=\columnwidth]{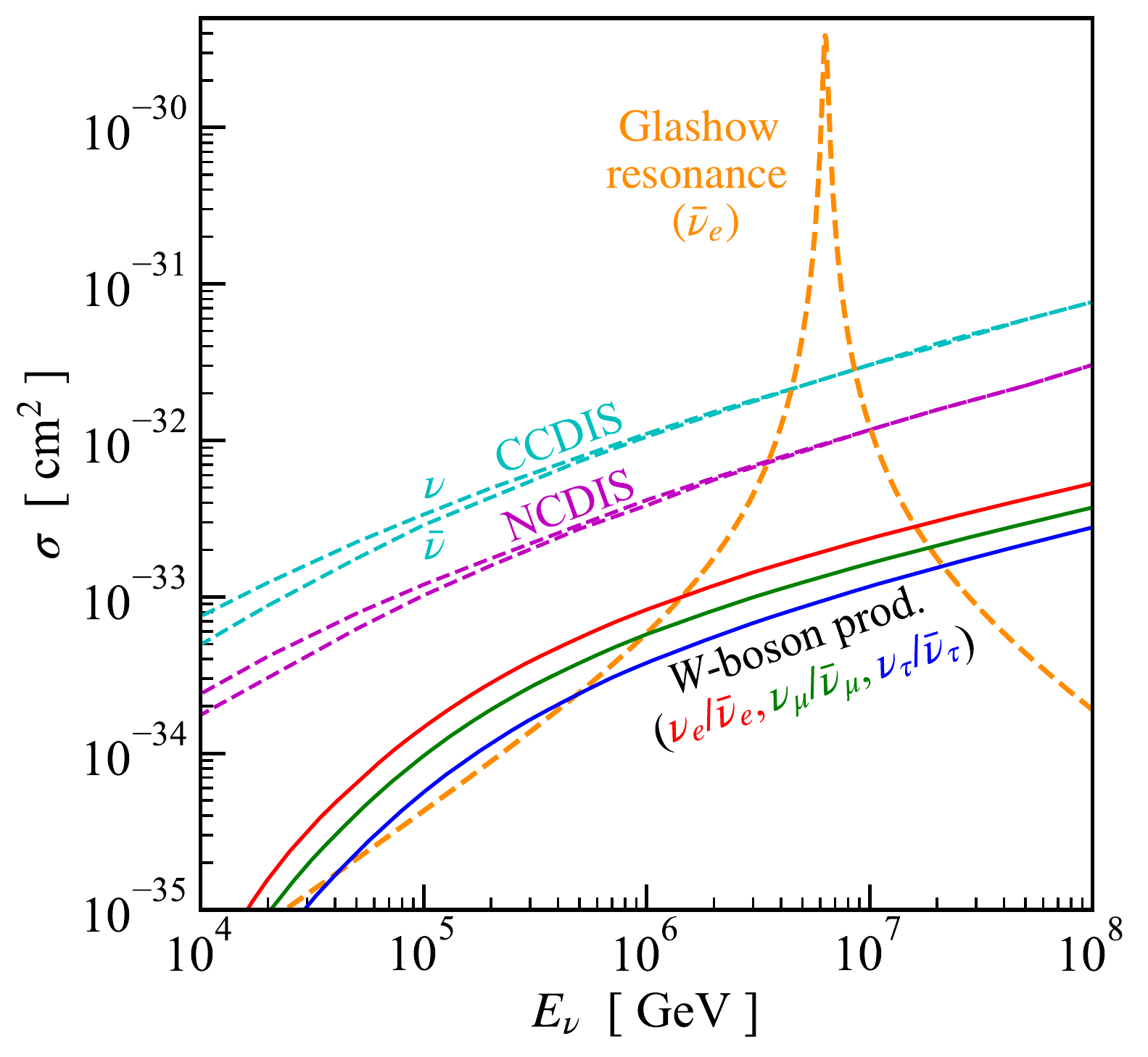}
\caption{Cross sections between neutrinos and $\ce{^16} \rm O$ for $W$-boson production~\cite{Zhou:2019vxt}, compared to those for CCDIS~\cite{CooperSarkar:2011pa}, NCDIS~\cite{CooperSarkar:2011pa}, and the Glashow resonance ($\bar{\nu}_e e^- \rightarrow W^-$, taking into account eight electrons)~\cite{Barger:2014iua}.}
\label{fig_sgmnuA}
\end{figure}


\section{$W$-boson production cross sections and implications}
\label{sec_xsec}

In this section, we briefly review the total cross section for $W$-boson production (Sec.~\ref{sec_xsec_total}; details are given in our companion paper~\cite{Zhou:2019vxt}), and present new calculations of the differential cross sections (Sec.~\ref{sec_xsec_diff}). Then we talk about the implications, including the cross section uncertainties (Sec.~\ref{sec_xsec_XsecUnc}) and the effects on neutrino attenuation in Earth (Sec.~\ref{sec_xsec_attenuation}).


\subsection{Review of the total cross sections}
\label{sec_xsec_total}

The nuclear production processes for on-shell $W$ bosons are
\begin{subequations}
\begin{align}
    \nu_\ell       + A & \rightarrow \ell^- + W^+ + A' \, , \\
    \bar{\nu}_\ell + A & \rightarrow \ell^+ + W^- + A' \, , 
\end{align}
\end{subequations}
where $A$ and $A'$ are the initial and final-state nuclei and $\ell$ is a charged lepton. The neutrino- and antineutrino-induced processes have the same total and differential cross sections, but there is flavor dependence. The coupling to the nucleus and its constituents is through a virtual photon, $\gamma^*$ (contributions from virtual $W$ and $Z$ bosons are only important for $E_\nu > 10^8$~GeV~\cite{Zhou:2019vxt}).  The process has a high threshold, $E_\nu \simeq 5 \times 10^3$~GeV, due to the large mass of the $W$ boson, though much lower than the threshold for the Glashow resonance, which peaks at $\simeq 6.3$~PeV. Above threshold, the leptonic decays of the $W$ boson (branching ratio $\simeq 11\%$ to each flavor) lead to the dominant contributions to trident production. 

\begin{figure*}[t!]
\includegraphics[width=\columnwidth]{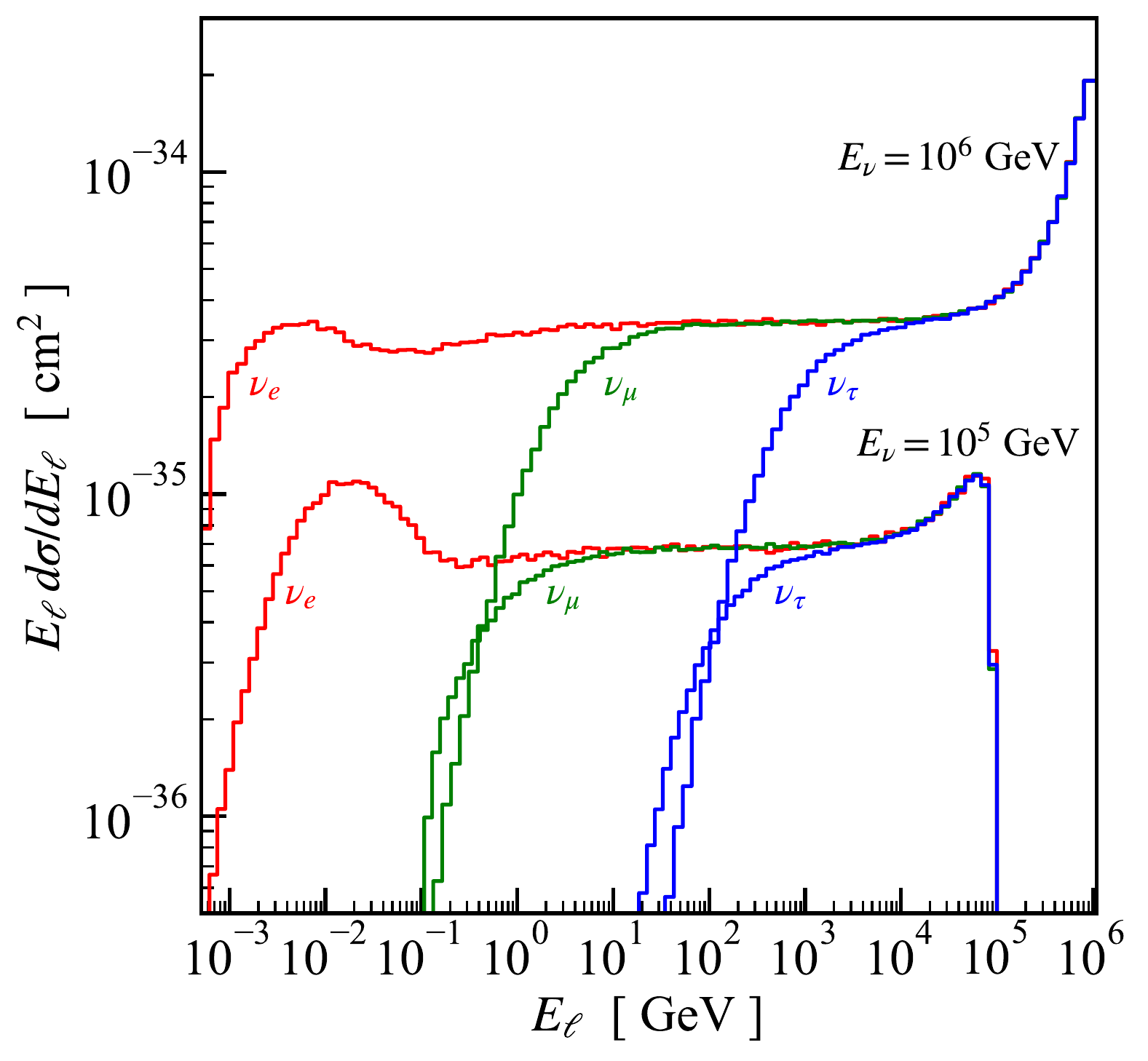}
\includegraphics[width=\columnwidth]{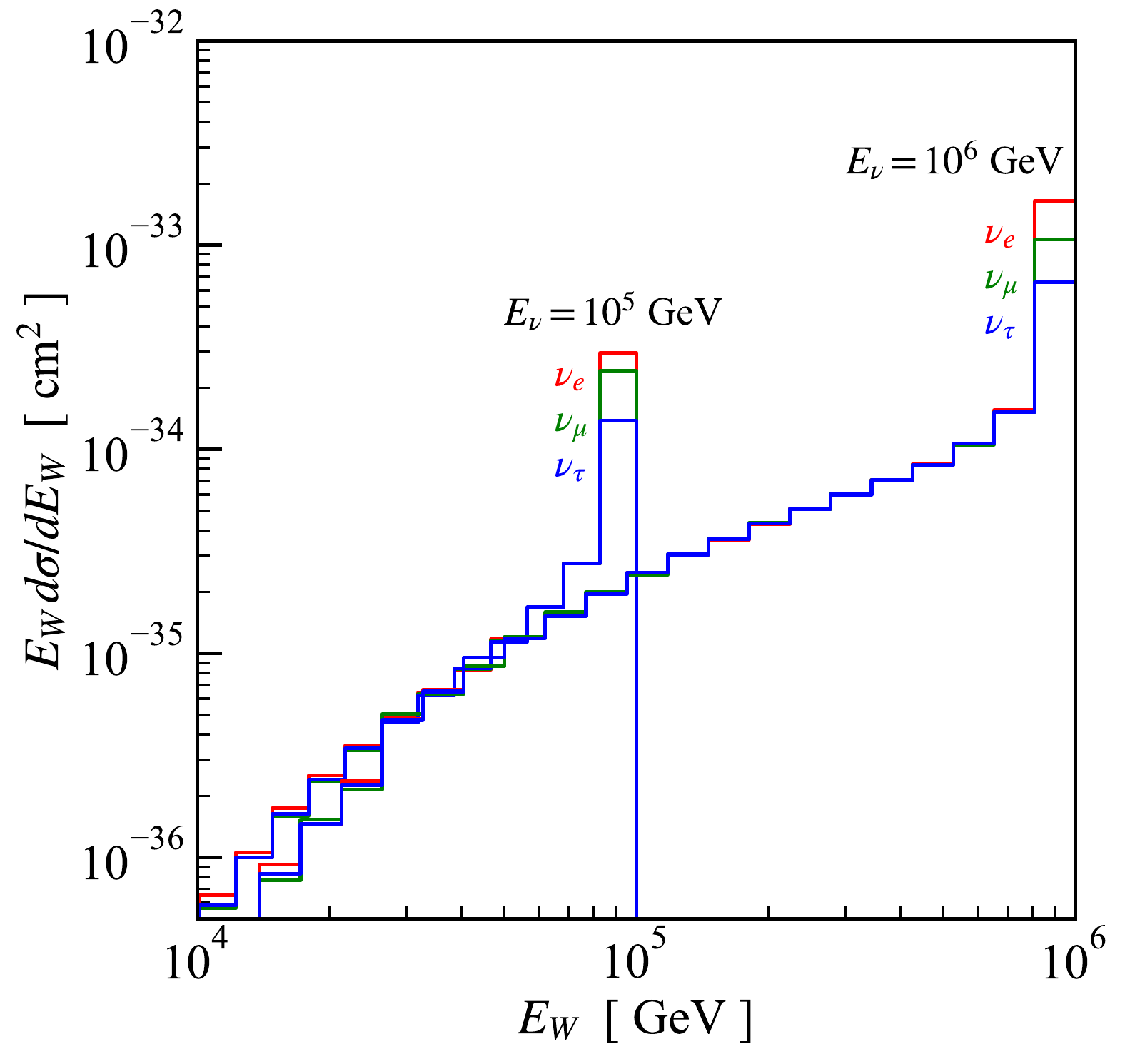}
\caption{
{\bf Left}: Differential cross sections for $W$-boson production in terms of the energy of the charged lepton, shown for each neutrino flavor and two typical energies ($E_\nu = 10^5$~GeV and $10^6$~GeV).  The $y$ axis is $E d\sigma/dE = (2.3)^{-1} d\sigma/d\log_{10}E$, matching the log scale on the x axis, so that relative heights of the curves at different energies faithfully show relative contributions to the total cross section.  {\bf Right:} Same, in terms of the energy of the $W$ boson.
}
\label{fig_sgmnuA_diffxsec}
\end{figure*}

The interactions happen in three different scattering regimes --- coherent, diffractive, and inelastic --- in which the virtual photon couples to the whole nucleus, a nucleon, and a quark, respectively. The corresponding cross sections are calculated separately and added to give the total cross section. For the coherent and diffractive regimes, we deal with the hadronic part in a complete way, which takes into account the photon virtuality, instead of using the equivalent photon approximation (as in, e.g., Refs.~\cite{Alikhanov:2014uja, Alikhanov:2015kla}). Moreover, in the diffractive regime, we include the Pauli-blocking effects that reduce the cross section~\cite{Bell:1996ms, Lovseth:1971vv, Ballett:2018uuc, Altmannshofer:2019zhy}. For the inelastic regime, we point out that there are two subprocesses: photon-initiated and quark-initiated. For the former, we use the up-to-date inelastic photon PDF of proton and neutron~\cite{Schmidt:2015zda, CT14web} and dynamical factorization and renormalization scales. For the latter, we do the first calculation and find that this sub-process can be neglected below $\simeq 10^8$~GeV.  {\it A key result is that our $W$-boson production cross section is smaller than that of previous work~\cite{Seckel:1997kk, Alikhanov:2014uja, Alikhanov:2015kla}.}

Figure~\ref{fig_sgmnuA} shows our $W$-boson production cross sections on $\ce{^16} \rm O$ for different neutrino flavors, along with other relevant processes. The width of the Glashow resonance is due to the intrinsic decay width of the $W$ boson.


\subsection{New results for the differential cross sections}
\label{sec_xsec_diff}

For the differential cross sections, the most relevant results to detection are the energy distributions of the charged lepton ($E_\ell$) and the $W$ boson ($E_W$). The energy that goes to the hadronic part is negligible (see next paragraph). As above, the differential cross sections are calculated separately for the three regimes and summed. For the coherent and diffractive regimes, the phase-space variables we chose to calculate the total cross section in Ref.~\cite{Zhou:2019vxt} are not directly related to the energies of the final states, so some transformations are needed; see Appendix~\ref{app_diffxsec_ElEW}. For the inelastic regime, following Ref.~\cite{Zhou:2019vxt}, we use {\tt MadGraph (v2.6.4)}~\cite{Alwall:2014hca} and analyze the event distributions in terms of the relevant quantities.

The energy that goes to the hadronic part, $\Delta E_h = Q^2/2 m_h$ (Appendix~\ref{app_diffxsec_ElEW}), is negligible compared to the detection threshold, which is $\sim 100$~GeV for showers in IceCube.  Here $Q^2 \equiv -q^2$ is the photon virtuality; the hadronic mass, $m_h$, is the nuclear mass in the coherent regime and the nucleon mass in the diffractive and inelastic regimes. For the coherent and diffractive regimes, the corresponding nuclear and nucleon form factors are highly suppressed above $Q \sim 0.1$~GeV and $Q \sim 1$~GeV, respectively, which leads to $\Delta E_h \lesssim (0.1)^2/2/16 \simeq 0.0003$~GeV and $\Delta E_h \lesssim (1)^2/2/1 \simeq 0.5$~GeV. For the inelastic regime, although $Q^2$ could be much larger, the cross section is still dominated by the low-$Q^2$ region ($Q^2 \lesssim 10$~GeV$^2$, i.e., $\Delta E_h \lesssim 10/2/1 \simeq 5$~GeV) because the nonperturbative part of the inelastic photon PDF~\cite{Schmidt:2015zda, Manohar:2016nzj} dominates the cross section (see Sec.~V.B of Ref.~\cite{Zhou:2019vxt}). {\it Above is very different the DIS, in which the energy transferred to the nucleus is $25\% E_\nu$ on average~\cite{Gandhi:1995tf, Gandhi:1998ri, CooperSarkar:2011pa, Connolly:2011vc}.}

Therefore, $E_\nu \simeq E_\ell + E_W$ is an excellent approximation for the coherent and diffractive regimes and a good approximation for the inelastic regime. We checked this through the distribution of the sum of $E_W$ and $E_\ell$ , finding that this is nearly a delta function at $E_\nu$.

Figure~\ref{fig_sgmnuA_diffxsec} shows the differential cross sections for the charged lepton (left) and $W$ boson (right), for each neutrino flavor and two typical energies, $E_\nu = 10^5$ and $10^6$~GeV, summed over the contributions from all three scattering regimes. For the charged lepton, the differential cross section is relatively flat when plotted as $E d\sigma/dE$, which means that no specific energy range is particularly favored; the narrow bump near $E_\nu$ does not contribute significantly to the total cross section. For the $W$ boson, the differential cross section favors the highest possible energy, $E_W \simeq E_\nu - m_\ell$. The differences between different flavors are due only to the charged lepton mass, $m_\ell$, which sets the lower limit of the distribution. Therefore, when $E_\ell \gg m_\ell$, the results for different flavors converge. This induces the opposite feature for the distribution of $W$ boson, where the results for different flavors converge at lower energies.

\begin{figure}[h]
\includegraphics[width=\columnwidth]{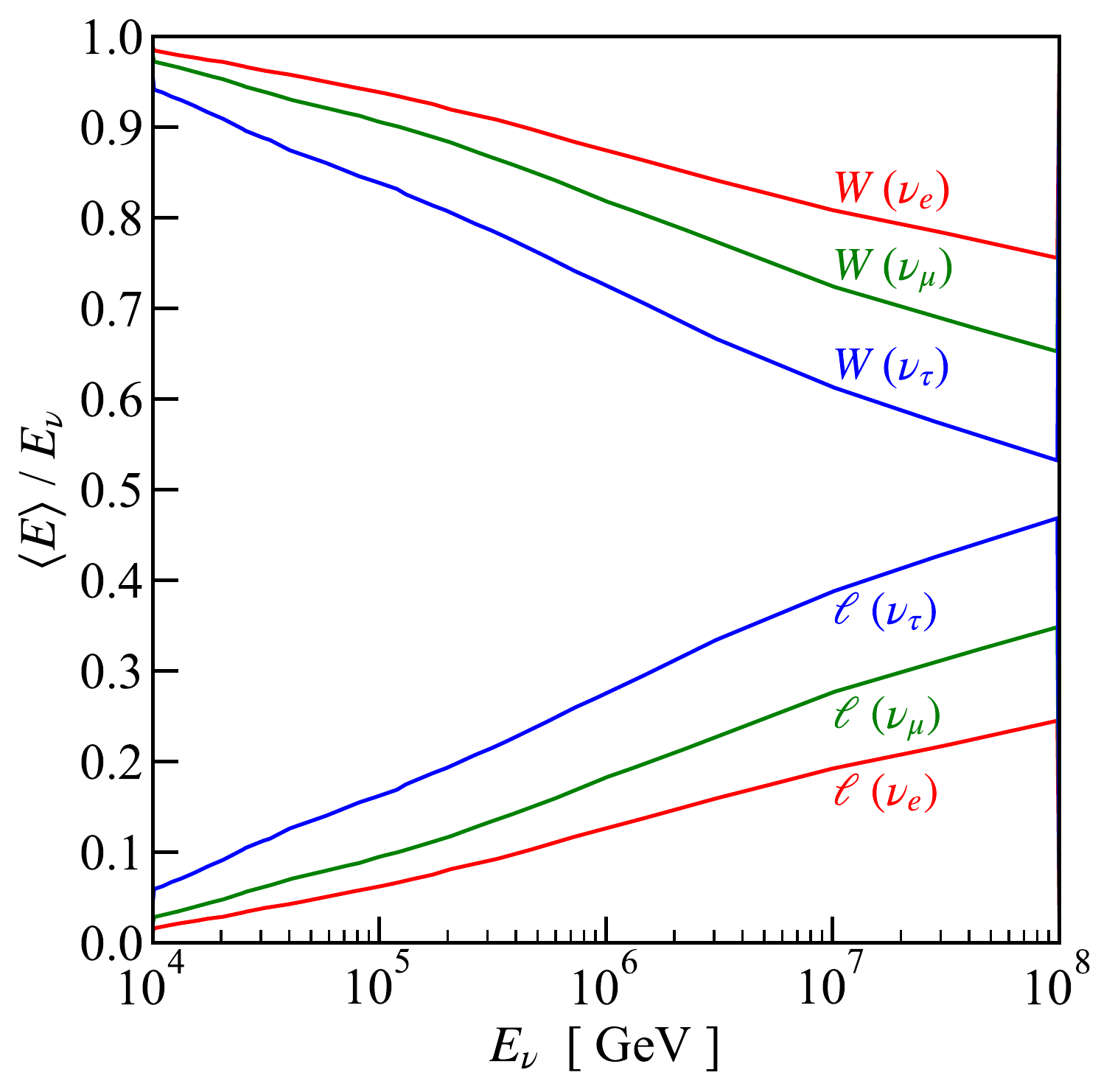}
\caption{
Average energy of the charged lepton ($\ell$) and $W$ boson, divided by $E_\nu$, for each neutrino flavor. 
}
\label{fig_sgmnuA_Eavgs}
\end{figure}

Figure~\ref{fig_sgmnuA_Eavgs} shows the average energy for the charged lepton $\ell$ and the $W$ boson for each flavor of initial neutrino.  This is calculated as 
\begin{equation}
    \frac{ \left< E \right> }{E_\nu} 
    = \frac{1}{E_\nu} \frac{ \int dE\, {E \frac{d\sigma}{d E}(E, E_\nu) } }{ \sigma(E_\nu) } \,,
\end{equation}
where $E = E_\ell$ or $E_W$.  As can be expected from Fig.~\ref{fig_sgmnuA_diffxsec}, the $W$ boson is typically much more energetic than the charged lepton, except for when $E_\nu$ is very large. In a crude approximation, at the main neutrino energies relevant to detection in IceCube or IceCube-Gen2, all the neutrino energy goes to the $W$ boson, with some dependence on neutrino flavor.  Below, we provide more careful calculations.

\subsection{Implication: Cross-section uncertainty}
\label{sec_xsec_XsecUnc}

\begin{figure}[t!]
\includegraphics[width=\columnwidth]{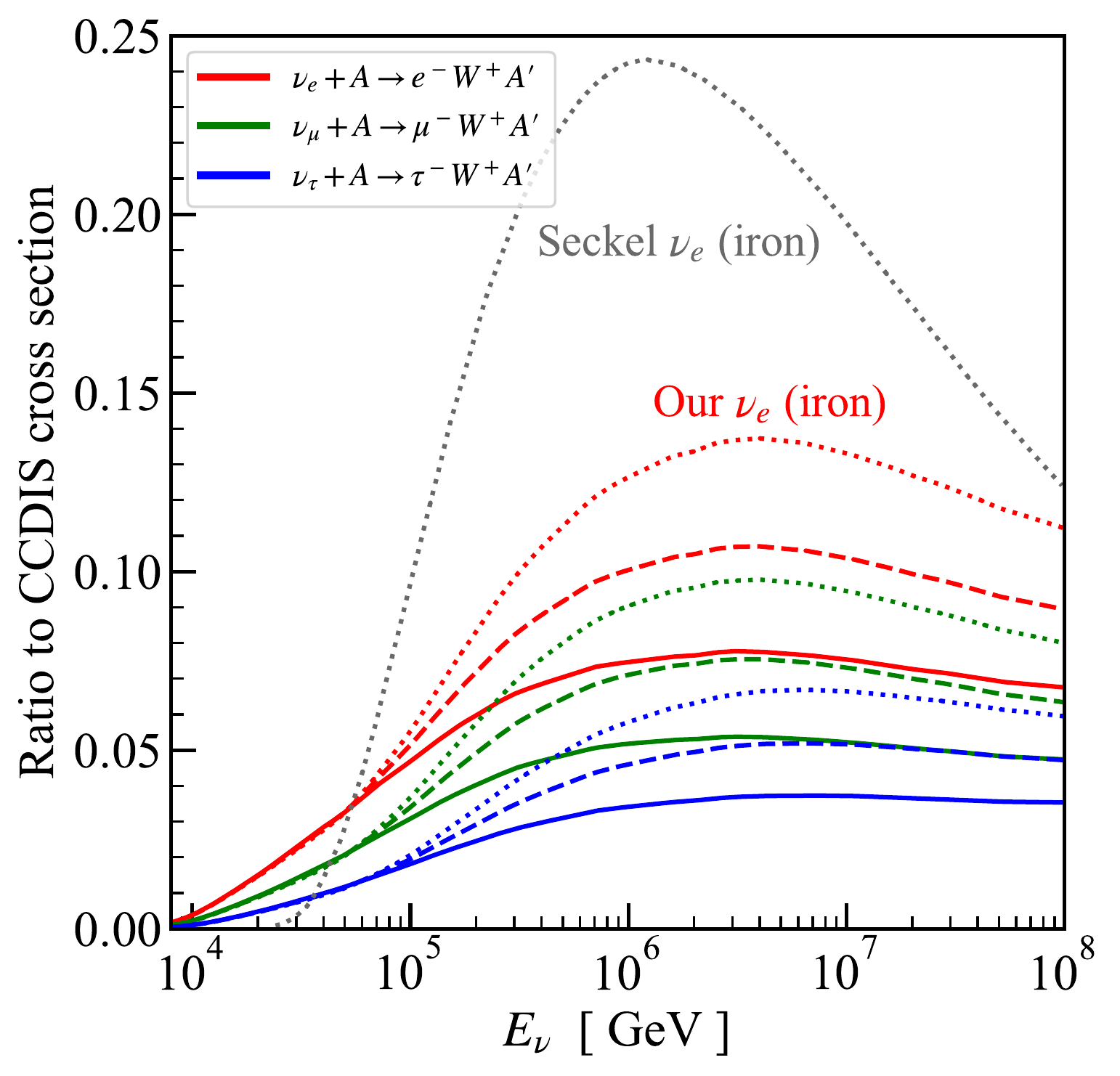}
\caption{Ratios of the $W$-boson production cross sections~\cite{Zhou:2019vxt} to those of CCDIS (($\nu + \bar{\nu}$)/2)~\cite{CooperSarkar:2011pa}.  Solid lines are for water/ice targets, dotted for iron targets, and dashed for Earth's average composition.  Color assignments are noted in the legend.  For comparison, we also show the $\nu_e$ (iron) result of Seckel~\cite{Seckel:1997kk}, which is much larger than ours.}
\label{fig_ratios}
\end{figure}

Figure~\ref{fig_ratios} shows the ratios of the $W$-boson production to the neutrino CCDIS cross sections ($( \sigma^{\rm CCDIS}_\nu + \sigma^{\rm CCDIS}_{\bar{\nu}} )/2$)~\cite{CooperSarkar:2011pa}.  We neglect the NCDIS cross section because it is smaller (see Fig.~\ref{fig_sgmnuA}) and because the energy deposition is only $\simeq 0.25 E_\nu$, which suppresses its importance~\cite{Beacom:2004jb}.  What we show is most relevant for detection. For a water/ice target, the maximum ratios of $W$-boson production to CCDIS~\cite{CooperSarkar:2011pa} are $\simeq 7.5\%$ ($\nu_e$), $\simeq 5\%$ ($\nu_\mu$), and $\simeq 3.5\%$ ($\nu_\tau$).  For an iron target or Earth's average composition, the maximum ratios are $14\%$/$11\%$ ($\nu_e$), $10\%$/$7.5\%$ ($\nu_\mu$), and $7\%$/$5\%$ ($\nu_\tau$).  This is more relevant to propagation (affected by both CCDIS and NCDIS) than detection (dominated by CCDIS); see Sec.~\ref{sec_xsec_attenuation} for details. The larger the charge number of a nucleus, the larger the ratio is.  The coherent component is $\propto Z^2$, while the diffractive and inelastic components are $\propto Z$, the same as for DIS.
{\it As noted, our results are significantly smaller than those of Seckel~\cite{Seckel:1997kk}.}

For the CCDIS cross section, the claimed uncertainties (from the parton-distribution functions) in $10^4$--$10^8$~GeV are 1.5--4.5\% in Ref.~\cite{CooperSarkar:2011pa} and 1--6\% in Ref.~\cite{Connolly:2011vc} (see also Refs.~\cite{Chen:2013dza, Bertone:2018dse}).  The impact of $W$-boson production is thus significant and thus should be included in future calculations of neutrino-nucleus cross sections.  Further, as IceCube has detected 60 events above 60~TeV (deposited energy) in the past 7.5~years~\cite{Stachurska:2019srh, Schneider:2019ayi}, this means that taking $W$-boson production into account is relevant for IceCube and essential for IceCube-Gen2, which would be about 10 times larger.  This is detailed in Sec.~\ref{sec_detect}.

For future calculations of the cross-section uncertainties, aiming to reach the few-percent scale, we note some other corrections that should be taken into account.  The DIS calculations are done at next-to-leading order in QCD, using corresponding parton distribution functions.  However, as far as we are aware, next-to-leading order electroweak corrections~\cite{Barlow:1978se, Petronzio:1978sg, Diener:2003ss, Arbuzov:2004zr, Diener:2005me} are not included.  At the highest energies, nonperturbative electroweak cascades~\cite{Berezinsky:2002hq} may be a significant effect.  Finally, there are other processes, such as tri-charged lepton production~\cite{Barish:1977bg, Benvenuti:1977zp, Hansl:1978cp, Barger:1978mv}, that become increasingly important at high energies. In addition, in Ref.~\cite{Klein:2019nbu}, Klein notes that going beyond assuming isoscalar nucleon targets and nuclear effects on the parton distribution function should also be considered.  

\subsection{Implication: Attenuation in Earth}
\label{sec_xsec_attenuation}

Starting in the TeV range, neutrinos may be significantly attenuated while passing through Earth.  (For a path along an Earth diameter, $\tau = 1$ at $E_\nu \simeq 40$ TeV.)  Attenuation depends on the total CCDIS + NCDIS cross section, $\sigma(E_\nu)$.  Taking into account NCDIS increases the cross section by a factor $\simeq 1.4$ compared to CCDIS only~\cite{Gandhi:1995tf, Gandhi:1998ri, CooperSarkar:2011pa, Connolly:2011vc}. We ignore neutrino regeneration because of the steeply falling neutrino spectra. 

The optical depth $\tau = C \sigma$, where $C(\cos \theta_z)$ is the target number column density integrated along the line of sight, which depends on the zenith angle, $\theta_z$.  We use Earth's average composition.  The flux is attenuated by a factor
\begin{equation}
    A = e^{ - \tau(E_\nu, \, \cos{\theta_z})}.
\end{equation}
The column density in the direction of the zenith angle $\theta_z$ is reasonably well known~\cite{DZIEWONSKI1981297}.  Though the change in the cross section due to $W$-boson production is not large, it affects the argument of the exponential.  For $W$-boson production, we calculate $\tau$ as the sum of results for the three regimes, taking into account that the targets in the coherent regime are nuclei, while in the diffractive and inelastic regimes they are nucleons.

\begin{figure}[t!]
\includegraphics[width=\columnwidth]{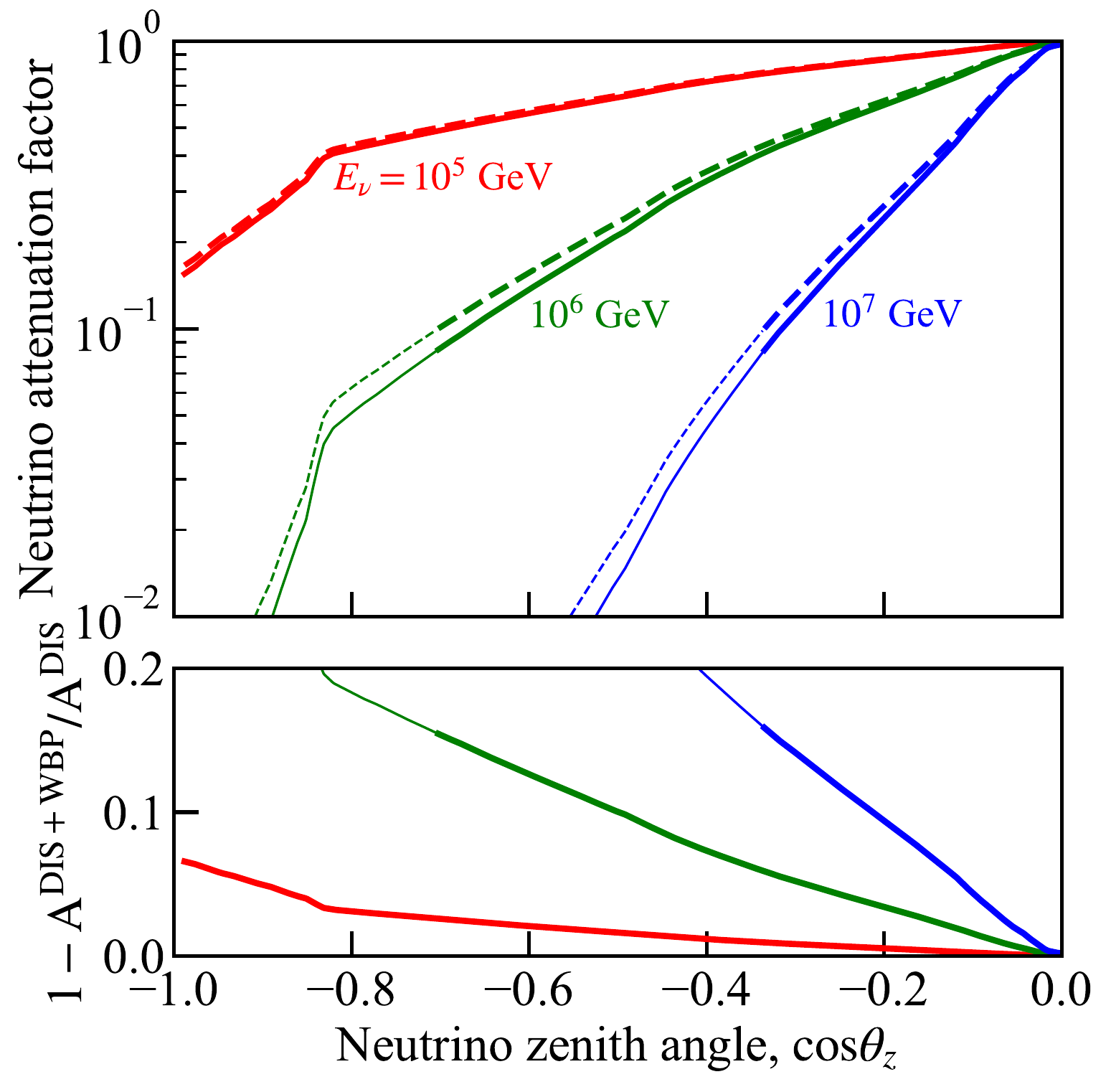}
\caption{
    {\bf Upper:} Neutrino attenuation factor, $e^{- \sigma C}$, for $\nu_e$ in Earth.  Dashed lines ($A^{\rm DIS}$) are for CCDIS and NCDIS without $W$-boson production.  Solid lines ($A^{\rm DIS+WBP}$) include $W$-boson production.  For attenuation factors below 0.1, the event rate is too low to use, which we denote by using thin lines.
{\bf Lower:} The relative change in the attenuation factor due to $W$-boson production.
}
\label{fig_attenuation}
\end{figure}

Figure~\ref{fig_attenuation} (upper panel) shows the neutrino attenuation factor without ($A^{\rm DIS}$) and with ($A^{\rm DIS+WBP}$) $W$-boson production.  For simplicity, we consider only $\nu_e$, which has the largest such cross section (see Fig.~\ref{fig_sgmnuA}).

Figure~\ref{fig_attenuation} (lower panel) shows the relative change to the attenuation factor, $1-A^{\rm DIS+WBP}/A^{\rm DIS}$.  For high energies and long paths through Earth, this can be quite large, even though the $W$-boson production cross section is small compared to the CCDIS cross section.  However, for $A$ that is too small, the event rate would be too low to matter; accordingly, we use thin lines for where $A \le 0.1$. Even avoiding these regions, the change in $A$ can be as large as $15\%$. This follows from
$ 
1-A^{\rm DIS+WBP}/A^{\rm DIS} = 1 - \exp(- C \sigma_{\rm WBP})  
\simeq C \sigma_{\rm DIS} \times \sigma_{\rm WBP}/\sigma_{\rm DIS} 
\simeq \tau \times \sigma_{\rm WBP}/\sigma_{\rm DIS} \,, 
$
which is the multiplication of the optical depth and the cross section ratio, which $\simeq 10\%/1.4 \simeq 7\%$ for $\nu_e$ (from Fig.~\ref{fig_ratios}).  The factor 1.4 roughly accounts for including NCDIS in addition to CCDIS.  For example, when $A = 0.1$, i.e., $\tau \simeq 2.3$, $ 1 - A^{\rm DIS+WBP}/A^{\rm DIS} \simeq 2.3 \times 7\% \simeq 15\%$, as above.

Interestingly, in IceCube's through-going muon analysis, there was an unknown $2\%$ deficit of straight up-going events, compared with their Monte Carlo simulation~\cite{Stettner:2019tok}. Taking $W$-boson production into account may explain this deficit.

The Earth attenuation effect allows a measurement of the neutrino-nucleus cross section at TeV--PeV energies in IceCube~\cite{Hooper:2002yq, Borriello:2007cs, Connolly:2011vc, Klein:2013xoa, Aartsen:2017kpd, Bustamante:2017xuy}.  The energy scales probed are far above those of laboratory experiments, for which the highest beam energies are $\simeq 350$ GeV~\cite{Seligman:1997fe, Tzanov:2005kr, Tanabashi:2018oca}.  In essence, the downgoing neutrino event rate depends on $\phi \sigma$, while the upgoing event rate depends on $\phi \sigma e^{-\tau}$, and taking a ratio cancels the flux and the detection cross section.

In present IceCube measurements of the cross section, the uncertainty is $\simeq 35\%$ when only the cross section normalization is checked (one wide energy bin for all data, plus assuming the shape of the standard model cross section)~\cite{Aartsen:2017kpd} or a factor of $\simeq 4$ when these assumptions are relaxed (several energy bins, no prior on the cross section shape)~\cite{Bustamante:2017xuy}. The energy ranges of both are comparable to where $W$-boson production is important. However, the measurement uncertainties will decrease. In addition, in Ref.~\cite{Aartsen:2017kpd} the ratio of the measured cross section to DIS prediction is $1.3 \pm 0.45$.  The central value would be about 0.1 smaller if the contribution from $W$-boson production were included.

Last, attenuation effects also lead to slightly altered flavor ratios because the $W$-boson production cross sections are flavor dependent.

\section{Detectability}
\label{sec_detect}

In this section, we calculate the detection prospects. We focus on $W$-boson production. The most important channels of trident production are a subset of $W$-boson production followed by leptonic decays~\cite{Zhou:2019vxt}. We first calculate the $W$-boson yields compared to those through the Glashow resonance (Sec.~\ref{sec_detect_Wyields}). Then, after a brief review of IceCube detection (Sec.~\ref{sec_detect_review}), we calculate the detectability of $W$-boson production from the shower spectrum (Sec.~\ref{sec_detect_showerspec}) and from unique signatures (Sec.~\ref{sec_detect_unique}).


\subsection{Larger $W$-boson yields than Glashow resonance}
\label{sec_detect_Wyields}

The Glashow resonance ($\bar{\nu}_e + e^- \rightarrow W^-$) is well known for producing on-shell $W$ bosons with a narrow feature in the cross section around $E_\nu = 6.3$ PeV.  The maximum cross section is $\sim 10^{-30}$~cm$^2$, a factor of about 100 larger than that of DIS.  Once the intrinsic and detector energy resolution are taken into account, the effects on the total event spectrum are less dramatic but still important.

Surprisingly, we find that on-shell $W$-boson production is actually dominated by neutrino-nucleus interactions where the coupling to the nucleus is through a photon.  The cross section is much smaller, but it involves all six neutrino flavors, and acts over a much wider energy range, in particular at lower energies, where the neutrino fluxes are much larger.

\begin{figure}[t!]
\includegraphics[width=\columnwidth]{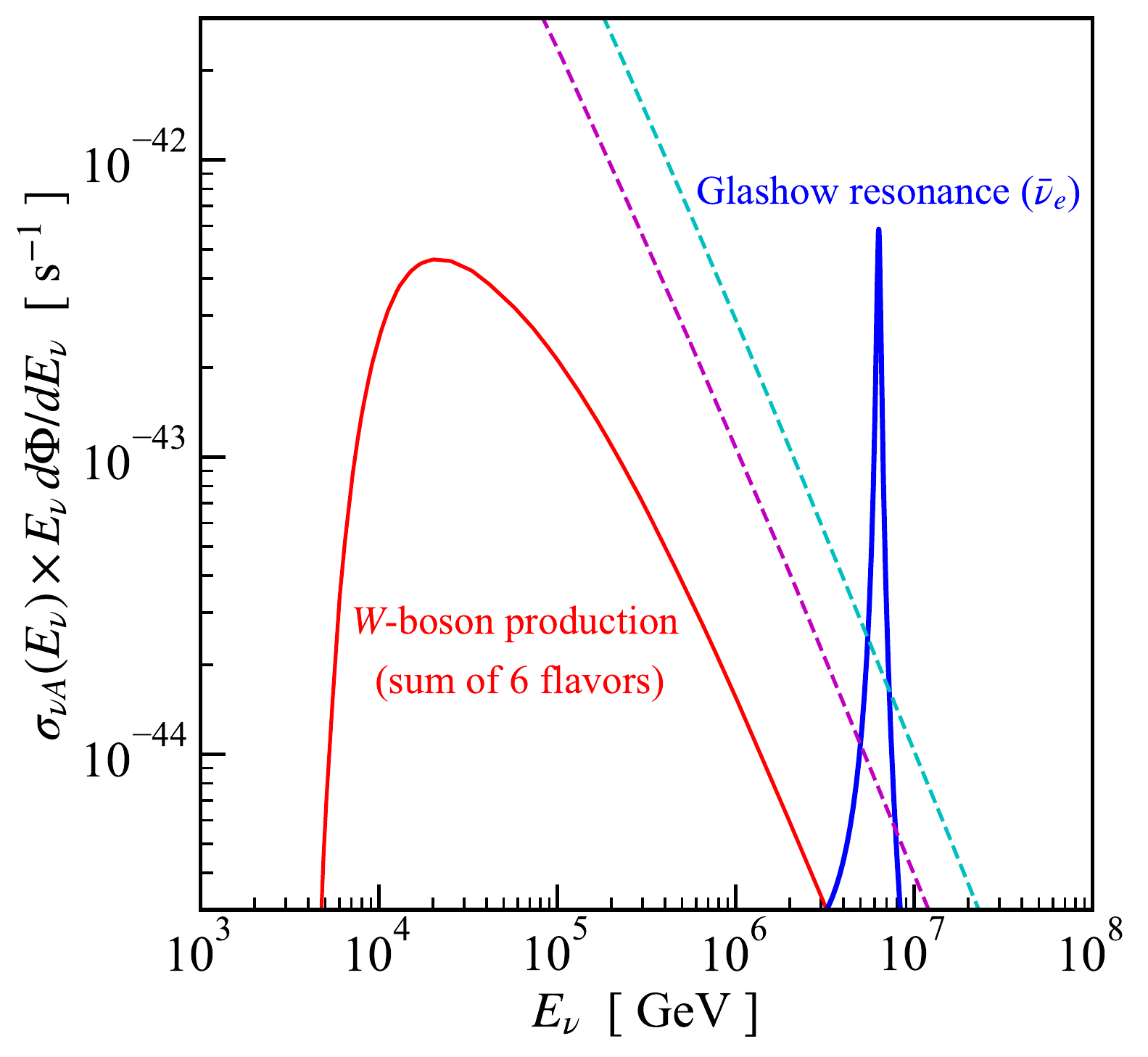}  
\caption{
Relative $W$-boson yields due to $W$-boson production ($\nu_l + A \rightarrow l + W + A'$) and the Glashow resonance ($\bar{\nu}_e + e^- \rightarrow W^-$).  We use $d\Phi/dE_\nu = (E_\nu/1 \rm \, GeV)^{-2.9}\, cm^{-2}\, s^{-1}\, GeV^{-1}$ with unit normalization.  The yield from $W$-boson production is $\simeq 20$ times that from the Glashow resonance, which can be seen by logarithmic integration of the peaks.  The CCDIS (cyan, dashed) and NCDIS (magenta, dashed) cases are shown for comparison, though they do not produce on-shell $W$ bosons.
}
\label{fig_Wyields}
\end{figure}

Figure~\ref{fig_Wyields} illustrates this. We multiply the cross sections by a power-law flux, $E_\nu \, d\Phi/dE_\nu = (E_\nu/1 \rm \, GeV)^{1 - \alpha}$ $\rm cm^{-2}\, s^{-1}\, GeV^{-1}$, with unit normalization, and plot results versus neutrino energy.  We use $\alpha = 2.9$, which matches the astrophysical neutrino spectrum from fitting IceCube data~\cite{Kopper:2017zzm, Schneider:2019ayi}.  (Below, we calculate more realistic expectations for detection.)  We multiply the flux by a factor $E_\nu$ so that the relative heights on the $y$ axis faithfully display the relative numbers of events per logarithmic energy bin.

The yield from neutrino-nucleus $W$-boson production is a factor of $\simeq 20$ times that for the Glashow resonance.  For $\alpha = 2.5$ and 2.0, the factor is $\simeq 3.5$ and $\simeq 0.5$ respectively.  {\it Therefore, for TeV--PeV neutrino observatories, neutrino-nucleus $W$-boson production is the dominant source of on-shell $W$ bosons unless the spectrum is very hard}.  This is a new and interesting physics point.  When it comes to detection, the $W$ bosons are not detected directly, due to their short lifetimes, and are instead detected by their decay products.


\begin{figure*}[t!]
\includegraphics[width=\columnwidth]{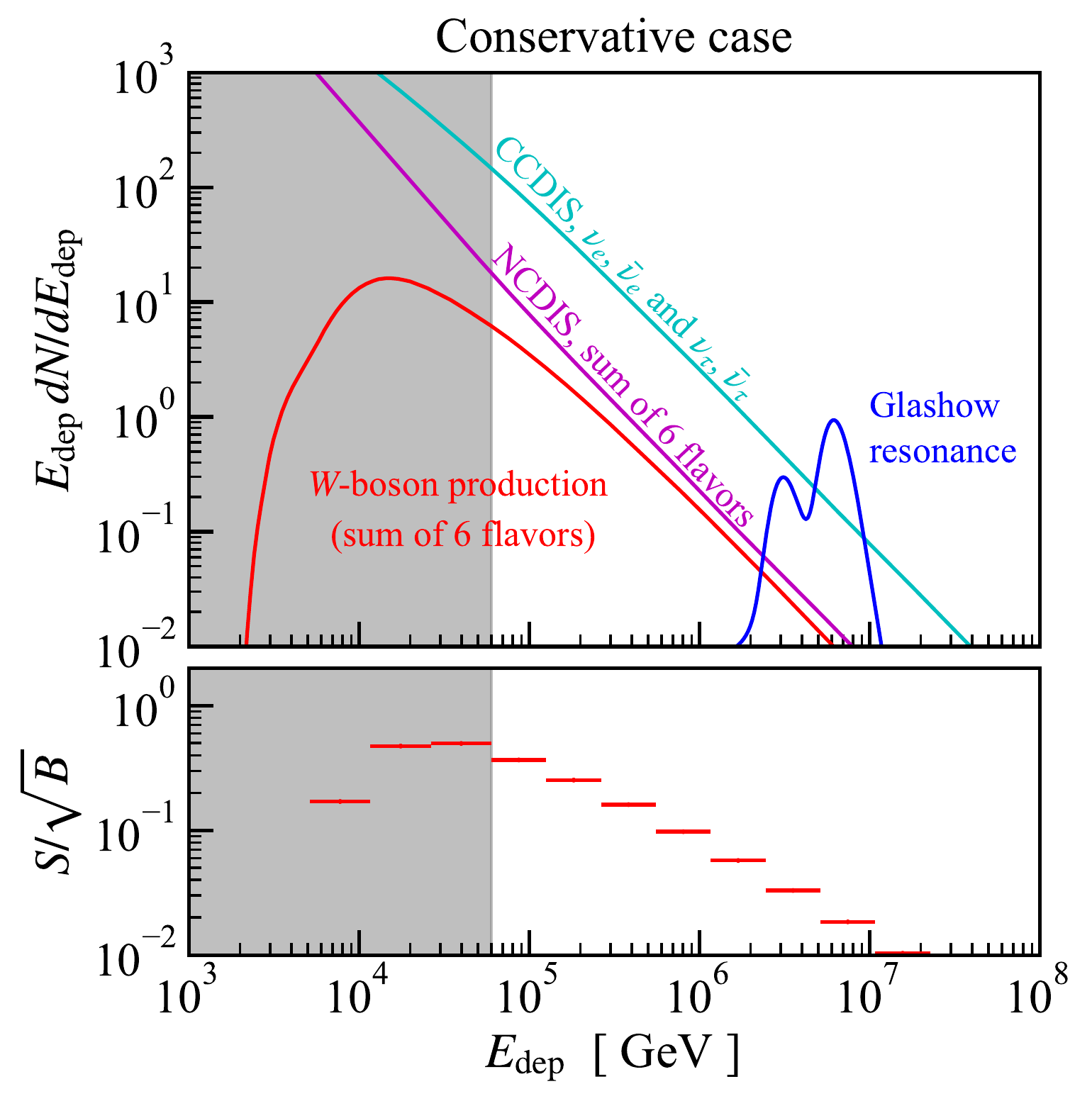}
\includegraphics[width=\columnwidth]{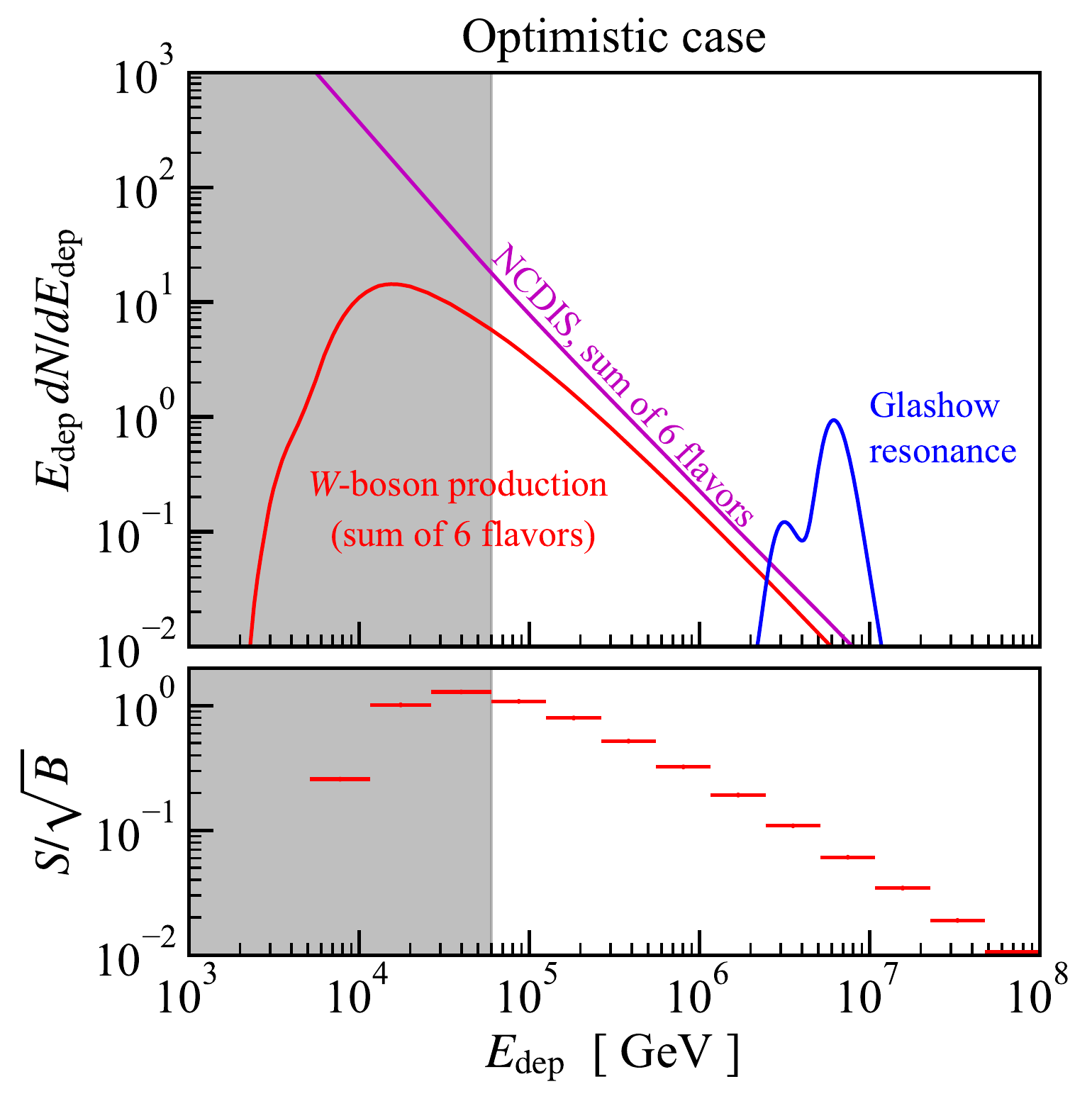}
\caption{
{\bf Left:} Shower spectrum ({\bf Upper}) and detection significance of $W$-boson production ({\bf Lower}) for the conservative case as regards identifying $W$-boson production events.  {\bf Right:} Same, but for the optimistic case.  The main difference between the two cases is the change with the CCDIS channel.  The shaded region below 60 TeV is below the IceCube threshold for cleanly identifying astrophysical neutrinos. We use 0.5~km$^3$ as the approximate fiducial volume of IceCube~\cite{Aartsen:2013jdh}, and assume 10 years of IceCube data.  For $E_{\rm dep} > 60$~TeV, where $W$-boson production contributes $\simeq 6$ shower events, the cumulative detection significance should be $\simeq 1.0\sigma$ for the conservative case and $3.2\sigma$ for the optimistic case.  See text for details.
}
\label{fig_event_spectrum}
\end{figure*}

\subsection{Review of detection in IceCube}
\label{sec_detect_review}

We briefly summarize the neutrino detection techniques used in IceCube and similar detectors~\cite{Aartsen:2013jdh, Aartsen:2015knd, Ahlers:2015lln, Li:2016kra}.  Neutrinos interact with nuclei and electrons, producing relativistic particles that emit Cherenkov light that is detected by photomultiplier tubes.

A $\nu_e$ CCDIS event produces an electron that carries most of the neutrino energy and hadrons that carry the remainder.  The electron initiates an electromagnetic shower of electrons, positrons, and gamma rays, with most of the Cherenkov emission coming from the most numerous low-energy but still relativistic charged particles.  The hadrons initiate a hadronic shower that consists primarily of pions.  The charged pions continue the hadronic shower, but the neutral pions decay promptly, feeding the electromagnetic shower.  The shower components induced by a $\nu_e$ event have high and comparable light yields, so that the total amount of Cherenkov light is proportional to $E_\nu$.  Because of the light scattering in ice, a shower looks like a large ($\sim 100$ m), round blob, even though the shower is a narrow cigar-shaped blob of length $\sim 10$ m.  For $\bar{\nu}_e$, the total and differential cross sections are slightly different, but the detection principles are the same.

For $\nu_\tau$ (and $\bar{\nu}_\tau$) CCDIS events, the results can be very similar to those for $\nu_e$ (and $\bar{\nu}_e$).  The $\tau$ decay produces a hadronic shower that is displaced in time and position, though these displacements start to become identifiable in IceCube only above $\sim 100$ TeV~\cite{Stachurska:2019wfb}.  At lower energies, $\nu_\tau$ events are nominally indistinguishable from $\nu_e$ events.  A way forward could be possible using muon and neutron echoes~\cite{Li:2016kra}.  For $\nu_\tau$ (and $\bar{\nu}_\tau$) events, the average deposited energy is $\simeq 20\%$ less than $E_\nu$ due to losses of neutrinos from $\tau$ leptonic decays.  In addition, 17\% of $\tau$ decays produce muon tracks, producing separable events.

All six flavors of neutrinos cause NCDIS events that also produce showers.  These appear identical to the other shower events above, though the energy deposition is typically only $\simeq 0.25 E_\nu$.  Because of the falling neutrino spectra, these events matter much less in the detection spectra~\cite{Beacom:2004jb}.

For $\nu_\mu$ (and $\bar{\nu}_\mu$) CCDIS events, the topologies are quite different because the muon range is so long, already $> 1$ km at $E_\mu \simeq 200$ GeV.  For events where the neutrino interaction is inside the detector (known as a contained-vertex or starting event), there is a hadronic shower and a long muon track, which itself produces small showers along its length.  Though the muon is not contained, its energy can be estimated from its energy-loss fluctuations, so that the neutrino energy can be estimated.  There can also be events where the neutrino interacts far outside the detector, and only the throughgoing muon is detected.  This enlarges the effective volume of the detector, but then only a lower limit on the neutrino energy can be set.

Shower events are especially important because of the ability to faithfully reconstruct the neutrino spectrum.  The shower spectrum can be estimated as~\cite{Kistler:2006hp, Laha:2013eev, Bustamante:2016ciw}
\begin{multline}
    E_{\rm dep} \frac{dN}{dE_{\rm dep}} = 2\pi \frac{\rho_{\rm ice} V_{\rm fid} N_{\rm A}}{18} T \int^{1}_{-1} d\cos\theta_z \times \\
E_\nu \frac{d \Phi}{dE_\nu}(E_\nu)\, \sigma(E_\nu)\, e^{-\tau(E_\nu,\, \cos\theta_z)} \,,
\label{eq_EventRate}
\end{multline}
where $E_{\rm dep}$ is the energy deposited in the detector from a shower, $\rho_{\rm ice} \simeq 0.92$~$\rm g\,cm^{-3}$ the density of ice, $V_{\rm fid} = 0.5$~km$^3$ is the approximate fiducial volume of IceCube~\cite{Aartsen:2013jdh}, $N_{\rm A}$ the Avogadro number (${\rho_{\rm ice} V_{\rm fid} N_{\rm A}}/{18} $ gives the number of water targets), and $T$ is the exposure time. For the neutrino flux, $\frac{d \Phi}{dE_\nu}$, we use that of Ref.~\cite{Kopper:2017zzm}, which includes both the atmospheric (dominated by $\nu_\mu$ and $\bar{\nu}_\mu$) and astrophysical neutrino fluxes.  The best-fit astrophysical flux, assuming 1:1:1 flavor ratios, is $(2.46 \pm 0.8) \times 10^{-18} (E/100\, {\rm TeV})^{-2.92}\, \rm GeV^{-1} \, cm^{-2} \, s^{-1} \, sr^{-1} $ for each flavor ($\nu + \bar{\nu}$), which is consistent with a more recent result~\cite{Schneider:2019ayi}. The $\sigma$ is the cross section between neutrino and water for different interaction channels. For CCDIS and NCDIS, we multiply the cross section on isoscalar nucleon targets~\cite{CooperSarkar:2011pa} by 18, the mass number for water.  For the Glashow resonance, we multiply the cross section on electrons by 10, the charge number for water. For the attenuation factor, $e^{-\tau(E_\nu,\, \cos\theta_z)}$, we use Ref.~\cite{Vincent:2017svp} with modification to include the cross section for $W$-boson production.  Once we obtain the shower spectra, we convolve them with a detector energy resolution of 15\%~\cite{Aartsen:2013vja}.


\subsection{Total shower detection spectrum}
\label{sec_detect_showerspec}

For the detection of $W$-boson production events, we consider two general scenarios.  In this subsection, we consider final states that contribute to the overall shower spectrum.  (The prospects for detection via track events are not favorable.)  In the next subsection, we consider unique final states that can be individually identified.  We focus on IceCube~\cite{Aartsen:2015knd}, the largest detector for TeV--PeV neutrinos.  Our results can be scaled to the proposed IceCube-Gen2, whose instrumented volume is expected to be 10 times that of IceCube~\cite{Blaufuss:2015muc}, and fiducial volume may be more, though with a higher energy threshold.

Table~\ref{tab_branchings} shows the complex possibilities for $W$-boson production events, including pure shower, track, and other unique signatures, depending on the decay modes of the $W$ boson and $\tau$ leptons. For the charged leptons from the initial interactions, their detectability depends on IceCube's trigger threshold ($\simeq 100$~GeV; note the analysis threshold of IceCube, $\simeq 1$~TeV, is less relevant because one would be searching for a lower-energy event in association with a higher-energy event)~\cite{Li:2016kra}.  Fig.~\ref{fig_sgmnuA_diffxsec} (left) is, up to an overall factor $2.3^{-1}$, the probability distribution in $\log_{10} E_\ell$, and it is roughly flat, with a median $\sim 100$~GeV. Therefore, we assume that half of the primary leptons are detectable and half not. In the ``Signatures'' column of Table~\ref{tab_branchings}, we use ``/'' to distinguish the two cases. (The decaying $W$'s are always detectable.)

First, we calculate the change to the overall shower spectrum in the conservative case where $\nu_\tau$ events appear as showers and where all showers are indistinguishable.  We ignore events with an energetic muon track: $\nu_\mu$ CCDIS, $\nu_\tau$ CCDIS with $\tau \rightarrow \mu$, half of $\nu_\mu$-induced $W$-boson production, $\nu_e$- and $\nu_\tau$-induced $W$-boson production with $W \rightarrow \mu$, $\tau \rightarrow \mu$, or $W \rightarrow \tau \rightarrow \mu$, and Glashow resonance events with $W \rightarrow \mu$ or $W \rightarrow \tau \rightarrow \mu$.

Second, we calculate for an optimistic case where $\nu_\tau$ CCDIS events are identifiable through a double-bang or double-pulse signature (above $\sim 100$ TeV, this is becoming realistic with current technology~\cite{Stachurska:2019wfb}), and where electromagnetic and hadronic showers can be separated using echo techniques (this is not yet possible with IceCube, but it may be with IceCube-Gen2).  Therefore, for CCDIS, we remove the remaining $\nu_e$- and $\nu_\tau$-induced channels. For the $W$-boson production, we remove the channels that give tracks, pure EM showers (EM means electromagnetic), and other unique signatures (see Table~\ref{tab_branchings}).

For $\nu_e$ CC, $E_{\rm dep} \simeq E_\nu$, as both the final state electron and hadrons produce showers. For $\nu_\tau$ CC, $E_{\rm dep} \simeq \left[ \left< y \right> + 0.7 (1-\left< y \right>) \right] E_\nu \simeq 0.8 E_\nu$, where $\left< y \right> \simeq 0.25$, for both CCDIS and NCDIS, is the average inelasticity, which is the fraction of neutrino energy transferred to the hadrons~\cite{Gandhi:1995tf}.  The factor 0.7 above is due to about 30\% of the energy is taken away by neutrinos from $\tau$ decay. For all-flavor NCDIS, $E_{\rm dep} \simeq \left< y \right> E_\nu \simeq 0.25 E_\nu$.  The ratios for CCDIS and NCDIS above are the similar to those used in Refs.~\cite{Laha:2013eev, Bustamante:2016ciw, Li:2016kra}.

For $W$-boson production, we use $E_{\rm dep} \simeq E_\nu$ if the $W$ decays hadronically and $E_{\rm dep} \simeq 0.5 E_\nu$ if the $W$ decays leptonically.  At the relevant energies, the $W$ boson takes nearly all of the neutrino energy; even when it does not, the approximations here are good for the total energy deposition, because the charged lepton ($e$ or $\tau$) from the initial interaction deposits most of its energy.  We make the same assumptions for $W$ bosons produced via the Glashow resonance.

Figure~\ref{fig_event_spectrum} shows the total shower spectrum for the conservative (left) and optimistic (right) cases for IceCube observations with $T = 10$ years (or 1 year of IceCube-Gen2~\cite{Blaufuss:2015muc}). $W$-boson production is subdominant, especially in the conservative case. The CCDIS events dominate, due to the large cross section and energy deposition. The NCDIS events are reduced in importance due the small energy deposition.  The two-peaks feature of the Glashow resonance is due to the leptonic and hadronic decays of $W$ bosons.

The lower panels show the detection significance. For each bin, this is calculated by the number of $W$-boson production events divided by the square root of CCDIS+NCDIS events.  The cumulative significance for detecting $W$-boson production, combining all the bins above $E_{\rm dep} = 60$~TeV, is $\simeq 1.0\sigma$ for the conservative case and $3.2\sigma$ for the optimistic case, for 10 years of IceCube observations.  The Glashow resonance events are not included because doing so would not appreciably affect the results.  

In summary, for $E_{\rm dep} > 60$~TeV and for 10 years of IceCube observations, $W$-boson production contributes $\simeq 6$ shower events, and the cumulative detection significance should be $\simeq 1.0\sigma$ for the conservative case and $3.2\sigma$ for the optimistic case. With 10 years of IceCube-Gen2, the counts would improve by a factor of $\simeq10$ and the significances by a factor $\sqrt{10} \simeq 3.2$.

\begin{table*}[t!]
    \caption{ \label{tab_branchings}
    Different final state particles, signatures, corresponding fractions, and counts in IceCube. The counts are for greater than 60~TeV deposited energy and 10 years of IceCube observations (or 1 year for IceCube-Gen2). The numbers in the ``Channel'' column are the maximal ratios to the CCDIS cross section with water/ice. The numbers in ``$W$ decay'' and ``$\tau$ decay'' columns are the branching ratios. For the ``Final state'' and ``$\tau$ decay'' columns, we omit the neutrinos; ``$h$'' means hadrons. The unique signatures are in {\bf boldface}. The ``/'' divides the cases in which the charged lepton from the initial interaction is undetectable or detectable, which, to a good approximation, is half-half. The ``Fractions'' column shows the fraction of that row relative to the whole channel, which is the multiplication between the branching ratios of $W$ and $\tau$ decay.
     }  
\medskip
\renewcommand{\arraystretch}{1.1} \centering 
\begin{tabular*}{0.98\textwidth}{c|c|c||c|c|c|c}
\hline
    Channel & $W$ decay & \begin{tabular}[c]{@{}c@{}} Final \\ state \end{tabular} & $\tau$ decay & Signature & Fraction & Counts \\
\hline \hline

    \multirow{6}{*}{    
    {  \begin{tabular}[c]{@{}c@{}} $\nu_e \rightarrow e W$ \\  (7.5\% rel. \\ to CCDIS) \end{tabular} }    
        }   
    & $e \nu_e$, 11\%                         & e e                         &       & {\bf Pure EM shower}                              & 11\%   & 0.34 \\
    & $\mu \nu_\mu$, 11\%	              &	e $\mu$	                    &       & {\bf Track without}/with {\bf shower}             & 11\%   & 0.34	\\

\hhline{~------}
    & \multirow{3}{*}{$\tau \nu_\tau$, 11\%} & \multirow{3}{*}{e $\tau$} &  $e$, 18\%  & {\bf Pure EM shower}                              & 2.0\% & 0.06	    \\
    &                                        &                           & $\mu$, 17\%  & {\bf Track without}/with {\bf (displaced) shower} & 1.9\% & 0.06\\
    &                                        &                           &   $h$, 65\%  & Shower                                            & 7.2\% & 0.22  \\
\hhline{~------}

    & $q \bar{q}$, 67\%	                    &	e $h$	                    &		&	Shower                                          & 67\%	& 2.08	\\ 
\hline \hline 
\multirow{8}{*}{
    {  \begin{tabular}[c]{@{}c@{}} $\nu_\mu \rightarrow \mu W$ \\ (5.0\% rel. \\ to CCDIS) \end{tabular} }}
    & $e \nu_e$, 11\%                        & $\mu$ e                       &               & {\bf Pure EM shower}/Track with shower    & 11\%	& 0.56	\\
    & $\mu \nu_\mu$, 11\%	                 & $\mu$ $\mu$                   &               & {\bf Single/Double tracks without shower} & 11\%	& 0.56 \\

\hhline{~------}
    &\multirow{3}{*}{$\tau \nu_\tau$, 11\%}  & \multirow{3}{*}{$\mu$ $\tau$} & $e$, 18\% & {\bf Pure EM shower/Track with (displaced) shower} & 2.0\% & 0.10  \\
    &                                        &                               & $\mu$, 17\% & {\bf Single/Double tracks without shower}        & 1.9\% & 0.10 \\
    &                                        &                               &  $h$, 65\% &  Shower/{\bf Shower with (displaced) track}      & 7.2\% & 0.36 \\
\hhline{~------}

    &	$q \bar{q}$, 67\%	                 &	      $\mu$ $h$              &              &  Shower/Shower with track&	67\%	& 3.41 \\

\hline \hline
\multirow{12}{*}{
    {  \begin{tabular}[c]{@{}c@{}} $\nu_\tau \rightarrow \tau W$ \\ (3.5\% rel. \\ to CCDIS) \end{tabular} } }	
        & \multirow{3}{*}{$e \nu_e$, 11\%}   &  \multirow{3}{*}{$\tau$ e}    & $e$, 18\%    & {\bf Pure EM shower}                               & 2.0\% &  0.02 \\
        &                                    &                               & $\mu$, 17\%  & {\bf Pure EM shower/Track with (displaced) shower} & 1.9\% &  0.02 \\
        &                                    &                               & $h$, 65\%    & {\bf Pure EM shower}/Shower                        & 7.2\% &  0.09 \\

\hhline{~------}
        &\multirow{2}{*}{$\mu \nu_\mu$, 11\%} & \multirow{2}{*}{$\tau$ $\mu$} & $\mu$, 17\% &	{\bf Single/Double tracks without shower}       & 1.9\% & 0.02 \\
        &                                     &                               & e or $h$, 83\% & {\bf Track without shower/with (displaced) shower} & 9.1\% & 0.11 \\

\hhline{~------}
        & \multirow{4}{*}{$\tau \nu_\tau$, 11\%} & \multirow{4}{*}{$\tau$ $\tau$}	& $e$ $e$, 3\%  & {\bf Pure EM shower}                                & 0.4\%	& 0.004	\\
        &                                          &                               	& $\mu$ $\mu$, 3\% & {\bf Single/Double tracks without shower}        & 0.3\%	& 0.004	\\
        &		                                   &                            	& $\mu$ e/$h$, 29\% & {\bf Track without shower/with (displaced) shower} & 3.1\%	& 0.04	\\
        &		                                   &                            	& $h$ $h$/$e$, 65\% &  Shower/Double bang                                & 7.2\%	& 0.09 \\
\hhline{~------}

    &	\multirow{2}{*}{$q \bar{q}$, 67\%}	&	\multirow{2}{*}{$\tau$ $h$}	& $e$ or $h$, 83\%    & Shower                   &	56\%	& 0.69 \\
    &	                                       	&	                           	& $\mu$, 17\%      & Shower/Shower with track &	11\%	& 0.14 \\
\hline \hline
 Total counts  &	                                  \multicolumn{4}{}     	                         	            &                          &	        & 9.44 \\

\hline 
\end{tabular*}
\end{table*}


\subsection{Unique signatures}
\label{sec_detect_unique}

Table~\ref{tab_branchings} also shows the unique signatures (in boldface) that $W$-boson production could give in IceCube, including fractions and counts. The events that give unique signatures are from leptonic decays following $W$-boson production, therefore they are also trident events.  The counts are calculated using Eq.~(\ref{eq_EventRate}). Some of them are background free compared to other standard model processes (DIS and the Glashow resonance). 
We focus on IceCube's high-energy analysis, so the counts are for greater than 60~TeV deposited energy and also for 10 years of IceCube observations. For lower-energy analysis of IceCube, such as the medium-energy starting events~\cite{Aartsen:2014muf} and the Enhanced Starting Track Event Selection~\cite{Silva:2019fnq}, these unique signatures could also exist.

There could be ``Double tracks'' signatures, with one muon track from the initial interaction ($\mu$ or $\tau \rightarrow \mu$) and the other from the decay of $W$ boson. See Table~\ref{tab_branchings} for all the contributing channels.
The calculated counts are $\simeq 0.34$. {\it Double tracks could also come from outside the detector, which would increase the counts.} Some of the ``Double tracks'' events may have separation angles ($\theta$) too small for them to be distinguished from a single track~\cite{Brown:1971xk}.  According to Refs.~\cite{Abbasi:2012kza, Kopper:2015rrp}, IceCube's resolution for double tracks is about 150~m. Therefore, for the ``Double tracks'' traveling 1~km, as long as the $\cos\theta \lesssim 0.99$, they can be separated. Moreover, the ``Double tracks'' would be a background for dimuon-type new physics searches~\cite{Albuquerque:2003mi, Albuquerque:2006am, Kopper:2015rrp}.

These ``Double tracks'' may also be identified because they are a subset of ``track without shower'' signatures. The no-shower feature is because the energy transferred to the hadronic part in $W$-boson production is mostly negligible (Sec.~\ref{sec_xsec_diff}). The ``track without shower'' signatures are mostly background free, because in CCDIS the energy transferred to hadronic part, for $E_\nu > 10^5$~GeV, is mostly above the IceCube threshold (100~GeV), according to the $d\sigma_{\rm CCDIS}/dy$ of Ref.~\cite{Gandhi:1995tf}. Moreover, these signatures could also come from other channels of $W$-boson production (see Table~\ref{tab_branchings} for details). The calculated total counts are $\simeq 0.96$. 

Interestingly, there is a ``track without shower'' candidate (Event 5) in the IceCube event list in Ref.~\cite{Aartsen:2013jdh} (arXiv version, page 15). It has no obvious shower activity, while all seven other track events have prominent showers at their starting points.
It is important to quantify the probability of this event coming from CCDIS. Event 5 deposited 71.4 TeV energy in IceCube, so the neutrino energy is $\sim 10^5$ GeV.  We can require that the hadronic energy be smaller than the full energy lost by muon in the initial 100 m of its path length. This is conservative, as it would double the average energy deposited in the first 100 m of the muon track compared to the second 100 m, which would be visible, unlike for Event 5.
According to Ref.~\cite{Chirkin:2004hz}, the corresponding energy loss of a $10^5$ GeV muon would be about 4 TeV. From $d\sigma_{\rm CCDIS}/dy$ of Ref.~\cite{Gandhi:1995tf} for $E_\nu = 10^5$ GeV, we can estimate the probability for having a hadronic energy smaller than a given value.  For 100 GeV, it is $\simeq 0.3\%$. For 4 TeV, it is $\simeq 10\%$.  Therefore, the probability for Event 5 being induced by CCDIS is small, even in conservative cases.

Moreover, ``track with shower'' events from $W$-boson production would also look different from CCDIS in terms of the inelasticity distribution.  For CCDIS, the dominant energy, $(1 - y) E_\nu$, goes to the track, with the smaller remainder, $y E_\nu$, going to the hadrons.  For those $W$-boson production events that are analyzed as CCDIS events with tracks, the non-track energy comes primarily from the $W$-boson decay, and this is typically much larger than the energy going to the track, in contrast to CCDIS.  Therefore, $W$-boson production events should be included in theoretical expectations of attempts to better measure the $\nu_\mu$ to $\bar{\nu}_\mu$ flux ratio and neutrino charged-current charm production~\cite{Aartsen:2018vez}. This may also provide a way to detect $W$-boson production.

There could also be ``Pure EM shower'' signatures, where EM means electromagnetic. The no-hadronic-shower feature is because the energy transferred to the hadronic part in $W$-boson production is mostly negligible (Sec.~\ref{sec_xsec_diff}). The major contributing channels are $W \rightarrow e$ following $W$-boson production induced by $\nu_e$ or by other flavors with the initial charged lepton below the trigger threshold (see Table~\ref{tab_branchings} for details). The calculated counts are $\simeq 0.82$. This signature could be background free with the echo technique~\cite{Li:2016kra}, same reason as for the ``Track without shower'' signature.

At last, there could also be the ``Track + displaced shower'' signatures (Table~\ref{tab_branchings}). The calculated counts are $\simeq 0.35$. However, due to the short lifetime of the $\tau$, it may be hard to identify them. The $\nu_\tau$ CC events with $\tau$ decay to muon will be a background. 

These unique signatures would help flavor identification. For example, the ``Double track'' and ``Track without shower'' signatures are dominated by $\nu_\mu$-induced $W$-boson production; the ``Pure EM shower'' are dominated by $\nu_e$-induced $W$-boson production, and the ``Track with displaced shower'' is dominated by $\nu_\tau$-induced $W$-boson production. The unique signatures are also backgrounds for exotic signals.


\section{Conclusions}
\label{sec_concl}

It is time to study the role of subdominant neutrino-nucleus interactions for TeV--PeV neutrinos. These cosmic neutrinos provide essential probes for neutrino astrophysics and physics, and the statistics of IceCube and especially IceCube-Gen2 demand greater precision in the theoretical predictions used to interpret the data.

The most important subdominant processes not yet taken into account are those where the interaction of a neutrino with a nucleus and its constituents is through a virtual photon, $\gamma^*$.  These processes are $W$-boson production ($\nu_l + A \rightarrow l + W + A'$) and trident production ($\nu + A \rightarrow \nu + \ell_1^- + \ell_2^+ + A'$).  In a companion paper, we present the more comprehensive and precise calculations of these cross sections at high energies~\cite{Zhou:2019vxt}.

In this paper, we study the phenomenological consequences of these processes at TeV--PeV energies for IceCube and related experiments.  We have five major results:

\setbox1=\hbox{1.}
\setlist{noitemsep, topsep=5pt, parsep=4pt, partopsep=0pt, leftmargin=0.4cm, labelindent=0cm, labelwidth=10cm, itemindent=*, labelsep=\dimexpr0.4cm-\wd1}

\begin{enumerate}

\item 
{\it These interactions are dominated by the production of on-shell $W$-bosons, which carry most of the neutrino energy.}  The most important trident channels are a subset of $W$-boson production followed by leptonic decays.  The energy partition follows from the calculation of the differential cross sections (Fig.~\ref{fig_sgmnuA_diffxsec}) and the average energies (Fig.~\ref{fig_sgmnuA_Eavgs}) of the final states.  The lepton takes a modest amount of the neutrino energy and, in stark contrast to DIS, the hadronic final state takes almost none.

\item
    {\it The cross section on water/iron can be as large as 7.5\%/14\% that of charged-current deep-inelastic scattering, much larger than the quoted uncertainty on the latter.}  From Fig.~\ref{fig_ratios}, the maximum ratios of $W$-boson production to CCDIS cross sections for water/ice targets are $\simeq 7.5\%$ ($\nu_e$), $\simeq 5\%$ ($\nu_\mu$), and $\simeq 3.5\%$ ($\nu_\tau$).  For iron targets, these are $\simeq 14\%$, $\simeq 10\%$, and $\simeq 7\%$.  These are significantly smaller than the early predictions of Seckel~\cite{Seckel:1997kk}. On the other hand, these ratios are much larger than the quoted uncertainties on the deep-inelastic scattering cross section for $E_\nu = 10^4$--$10^8$~GeV, which are 1.5--4.5\% in Ref.~\cite{CooperSarkar:2011pa} and 1--6\% in Ref.~\cite{Connolly:2011vc}.  We also point out other corrections to DIS that should be taken into account for future calculations.

\item 
{\it Attenuation in Earth is increased by as much as 15\% due to these cross sections (Fig.~\ref{fig_attenuation}). They are also an inseparable part of the measured neutrino cross section.}  Though the uncertainty of measured cross sections by IceCube is larger than the change in the cross section due to $W$-boson production~\cite{Aartsen:2017kpd, Bustamante:2017xuy}, the measured uncertainties will decrease.  In addition, in Ref.~\cite{Aartsen:2017kpd} the ratio of the measured cross section to DIS prediction is $1.3 \pm 0.45$. The central value would be about 0.1 smaller if the contribution from $W$-boson production were included.

\item
{\it $W$-boson production on nuclei exceeds that through the Glashow resonance on electrons by a factor of $\simeq 20$.}  From Fig.~\ref{fig_Wyields}, the former produces on-shell $W$ bosons $\simeq 20$ times more efficiently than the Glashow resonance if the neutrino spectrum index is 2.9, the nominal value from fitting IceCube data~\cite{Kopper:2017zzm, Schneider:2019ayi}. This point was not previously known.

\item
    {\it The primary signals are showers that will significantly affect the detection rate in IceCube-Gen2; a small fraction of events give unique signatures that may be detected sooner.}  The overall shower spectrum is changed by $W$-boson production.  Based on the calculations in Fig.~\ref{fig_event_spectrum}, we show that this could be detected with 10 years of IceCube data above 60 TeV with significance $1.0\sigma$ and $3.2\sigma$ for conservative and optimistic cases.  In 10 years of IceCube-Gen2, these would improve by a factor $\sqrt{10} \simeq 3.2$.  We also note unique signatures that may be identified sooner, including with IceCube. (Though not explored here, it would be interesting to consider their impact on detectors for ultra-high-energy neutrinos.)

\end{enumerate}

Since 2013, IceCube has opened the field of high-energy neutrino astronomy, probing neutrino-nucleus interactions well above 1 PeV, far beyond the reach of laboratory experiments.  Now, only six years later, it is becoming important to take into account subdominant neutrino-nucleus interactions.  This rapid progress hints at the discovery prospects of larger detectors, both for increased precision in probing astrophysics and the cross section as well as in searches for new physics.

\vspace{0.9cm}
\section*{Acknowledgments}
We are grateful for helpful discussions with Brian Batell, Tyce DeYoung, Francis Halzen, Matthew Kistler, Shirley Li, Pedro Machado, Olivier Mattelaer, Kenny Ng, Yuber Perez-Gonzalez, Ryan Plestid, Carsten Rott, Ibrahim Safa, Subir Sarkar, Juri Smirnov, Tianlu Yuan, Keping Xie, and especially Mauricio Bustamante, Spencer Klein, Sergio Palomares-Ruiz, Mary Hall Reno, and David Seckel.

We used {\tt FeynCalc}~\cite{Mertig:1990an, Shtabovenko:2016sxi} and {\tt MadGraph}~\cite{Alwall:2014hca} for some calculations. 

This work was supported by NSF grant PHY-1714479 to JFB. BZ was also supported in part by a University Fellowship from The Ohio State University.


\onecolumngrid

\newpage 

\appendix

\section{$d \sigma_{\nu A}/d E_\ell$ and $d\sigma_{\nu A}/dE_W$}
\label{app_diffxsec_ElEW}

In this section, we detail the calculations of the differential cross section for $W$-boson production in the coherent and diffractive regimes. For these two regimes, we use the formalism in Refs.~\cite{Czyz:1964zz, Ballett:2018uuc}, and need to deal with the phase space by ourselves. For the inelastic regime, the phase space is handled with MadGraph~\cite{Alwall:2014hca}.

In our companion paper~\cite{Zhou:2019vxt}, we use the center-of-momentum (CM) frame between the neutrino and the virtual photon, which is the most convenient for calculating the total cross section. In this frame, the 4-momentum can be easily written as
\begin{subequations}
\begin{align}
    k_1 &= \left( \frac{s+Q^2}{2 \sqrt{s}},0,0,\frac{s+Q^2}{2 \sqrt{s}} \right) \,,\\
    q   &= \left( \frac{s-Q^2}{2 \sqrt{s}},0,0,-\frac{s+Q^2}{2 \sqrt{s}} \right) \,,\\
    p_1 &= (E_1, 0, -p \sin\theta, -p \cos\theta) \,, \\
    p_2 &= (E_2, 0, -p \sin\theta, -p \cos\theta) \,, 
\end{align}
\label{eq_app_4pCM}
\end{subequations}
where $k_1$, $q$, $p_1$, and $p_2$ are the 4-momenta of the neutrino, virtual photon, charged lepton, and $W$ boson, respectively, $s = (k_1+q)^2$, $Q^2 = -q^2$, and $p = \left\{\left[s-(m_l+m_W)^2\right] \left[ s-(m_l-m_W)^2 \right] \right\}^{1/2}/2\sqrt{s}$. 

However, to get the differential cross sections, $d \sigma_{\nu A}/d E_\ell$ and $d\sigma_{\nu A}/dE_W$, we should transform to the lab frame (nucleus-rest frame), in which
\begin{subequations}
\begin{align}
    k_1  &= \left( E_\nu, 0, 0, E_\nu \right) \,,\\
      q  &= \left( q_0, 0, q' \sin\theta_q, q' \cos\theta_q \right) \,, \\
      P  &= \left( m_h, 0, 0, 0 \right) \,, \\
      P' &= \left( m_h-q_0, 0, -q' \sin\theta_q, -q' \cos\theta_q \right) \,, 
\end{align}
\end{subequations}
where $P$ and $P'$ are the 4-momenta of initial and final nucleus or nucleon, and $m_h$ is its mass. It can be shown that $q_0$ = $-Q^2/2 m_h $, and $q' = \sqrt{ (Q^2/2m_h)^2 + Q^2}$. {\it Note that the energy transferred to the hadronic part is $\Delta E_h = -q_0 = Q^2/2 m_h$, which is small. }

The transformation from the neutrino-virtual photon CM frame to the lab frame can be done by two boosts first along the $z$ axis and then the $y$ axis (with Lorentz factors $\gamma_z$ and $\gamma_y$), and then a rotation by $x$ axis ($\cos{\omega_q}$). It can be shown that,

\begin{subequations}
\begin{align}
    \gamma_z &=  \frac{(s+Q^2) (2 E_\nu m_h-Q^2)}{ \sqrt{8 E_\nu m_h s Q^2 (2 E_\nu m_h-s-Q^2)}} \,,\\
    \gamma_y &=  \frac{1}{(s+Q^2)} \sqrt{\frac{2 E_\nu Q^2 (2 E_\nu m_h-s-Q^2)}{m_h}} \,,\\
    \cos{\omega_q} &= \frac{1}{\gamma_y}\,.
\end{align}
\end{subequations}

Therefore, the energy of the charged lepton in the lab frame, $E_\ell$, is,
\begin{equation}
\begin{split}
    E_\ell &= \frac{E_1 \left(2 E_\nu-\frac{Q^2}{m_h}\right)+p \cos\theta \left(E_\nu \left(2-\frac{4 s}{s+Q^2}\right)-\frac{Q^2}{m_h}\right)}{2 \sqrt{s}}+p \sin\theta \sqrt{\frac{4 E_\nu^2 Q^2}{(s+Q^2)^2}-\frac{2 E_\nu Q^2}{m_h (s+ Q^2)}-1} \\
    & = \frac{4 E_\nu m_h p Q^2 \cos\theta-\hat{s} (E_1 (Q^2-2 E_\nu m_h)+p \cos\theta (2 E_\nu m_h+Q^2))}{2 m_h \hat{s} \sqrt{\hat{s}-Q^2}}+p \sin\theta \sqrt{\frac{2 E_\nu Q^2 (2 E_\nu m_h-\hat{s})}{m_h \hat{s}^2}-1} \,,
\end{split}
\label{eq_app_El}
\end{equation}
in which the second step is just rewriting $s$ in terms of $\hat{s} = 2(k_1 \cdot q) = s+Q^2$, which is the phase-space variable for the cross-section calculation (see Eq.~(10) of Ref.~\cite{Zhou:2019vxt}). (In Eq.~(\ref{eq_app_4pCM}), we use $s$ to make the expressions more symmetric.)

Then, with some approximations,
\begin{equation}
\begin{split}
    E_\ell
    &\simeq \frac{E_\nu (E_1-p \cos\theta)}{\sqrt{\hat{s}}}+p \sin\theta \sqrt{\frac{4 E_\nu^2 Q^2}{\hat{s}^2}-\frac{2 E_\nu Q^2}{m_h \hat{s}}-1} \\
    &= \frac{E_\nu E_1}{\sqrt{\hat{s}}} - \frac{E_\nu p}{\sqrt{\hat{s}}}\cos\theta + p\sin\theta \frac{E_\nu Q}{\hat{s}} \left( \frac{Q^2}{m_h^2}+4 \right)^{1/2} \sin \theta_q \\
    &\simeq \frac{E_\nu}{\sqrt{\hat{s}}} \left( E_1 - p\cos\theta \right) \,.
\end{split}
\label{eq_app_El1}
\end{equation}
For the above steps, in a word, we basically ignore the terms with $Q^2$, motivated by the following. First, $Q$ is much smaller than the scale at which the interaction happens. Specifically, the $Q$ is highly suppressed above $\simeq 0.1$~GeV in the coherent regime (nuclear form factor) and $\simeq 1$~GeV in the diffractive regime (nucleon form factor), while $\sqrt{ \hat{s} } > m_W+m_\ell \simeq 80$~GeV and $E_\nu \gtrsim 4\times10^3$~GeV.
Second, the energy scale at which the $W$-boson production actually matters (ratio to CCDIS cross section is large; Fig.~\ref{fig_ratios}) is much higher than the threshold given above.

Therefore, following Eqs.~(10)--(11) of Ref.~\cite{Zhou:2019vxt}, 
\begin{equation}
\begin{split}
    \sigma_{\nu A} 
    & = \int_{ \hat{s}_{\rm min} }^{ \hat{s}_{\rm max} } d\hat{s} \int_{Q^2_{\rm min}(\hat{s}) }^{ Q^2_{\rm max}(\hat{s}) } dQ^2 \int_{-1}^{1} d\cos\theta \, \simeq
    \int_{ \hat{s}_{\rm min} }^{ \hat{s}_{\rm max} } d\hat{s} \int_{Q^2_{\rm min}(\hat{s}) }^{ Q^2_{\rm max}(\hat{s}) } dQ^2 \int_{E_{l,\rm min}(\hat{s})}^{E_{\ell,\rm max}(\hat{s})} \frac{\sqrt{\hat{s}}}{E_\nu p} dE_\ell \,  \\
    & \simeq \int_{E_{l,\rm min}}^{E_\ell, \rm max} \frac{\sqrt{\hat{s}}}{E_\nu p} dE_\ell \int_{ \hat{s}_{\rm min}(E_l) }^{ \hat{s}_{\rm max}(E_l) } d\hat{s} \int_{Q^2_{\rm min}(\hat{s})}^{Q^2_{\rm max}(\hat{s})} dQ^2 \,  \,,
\end{split}
\end{equation}
from which we can get ${d \sigma_{\nu A}}/{d E_\ell}$. In the equation above, the integrand is not shown.
Here $E_{\ell, \rm max} = E_\nu (E_1 + p)/\sqrt{\hat{s}}$ and $E_{\ell, \rm min} = {\rm \tt Max}\left\{ E_\nu (E_1 - p)/\sqrt{\hat{s}},\  m_\ell \right\} $ are the upper and lower limits of $E_\ell$, obtained from Eq.~(\ref{eq_app_El1}). 

The $d \sigma_{\nu A} / d E_W$ is obtained following a similar procedure.


\newpage
\twocolumngrid
\bibliography{references}
\end{document}